\documentclass{PoS}

\title{Jet Measurements In CMS}

\ShortTitle{Jet Measurements In CMS}

\author{\speaker{Sanmay Ganguly}\\
        Department Of High Energy Physics\\
        Tata Institute Of Fundamental Research\\
        Mumbai, India \\
        E-mail: \email{Sanmay.Ganguly@cern.ch}}

\abstract{
   A measurement of inclusive jet and dijet production cross sections is presented \cite{QCD}.
   Data from large hadron collider (LHC) proton-proton collisions at $\sqrt{s}=$ 7 TeV, corresponding to
   $4.67 fb^{-1}$ of integrated luminosity, have been collected with the compact muon solenoid (CMS) detector \cite{DETECTOR}.
   Jets are reconstructed with the anti-$k_T$ clustering algorithm with size parameter
   $R=0.7$, extending to rapidity $|y|=2.5$, transverse momentum $p_{T}=$ 2 TeV, and dijet
   invariant mass $M_{JJ}=$ 5 TeV. The measured cross sections are corrected for detector
   effects and compared to perturbative QCD predictions at next-to-leading order (NLO), corrected for
   non perturbative (NP) factors, using
   various sets of parton distribution functions. \\
   Determination Of Jet Energy Correction from $\sqrt{s}=$ 7 TeV CMS data is presented. The individual components are
   determined. The jet energy scale uncertainty factors are also shown.}

\FullConference{36th International Conference on High Energy Physics,\\
		July 4-11, 2012\\
		Melbourne, Australia}

\begin{document}

\section{Introduction}
At LHC proton-proton collision is taking place at very high center of mass energy $\sqrt{s}=$ 7 TeV.
At such energy scale, the fundamental interacting objets are partons, e.g. quarks, gluons etc. After the collision, the hard
 scattered partons undergoes fragmentation and then hadronization to form colorless bound states. These colorless objects deposit
 energy at dfferent components of the detector. To study the parton level dynamics, jet clustering algorithm is applied on
 each stable particles in an event, reconstructed from every individual component of the detector. \newline
 Jet study is important from both thoretical and experimental point of view. 
 It is also used to improve different parton shower model and higher order theory prediction. Measured cross section of jet
 processes are often used to estimate the number of background events for new physics studies, as well as constrain the parameters
 of parton distribution function.  

\section{Jet Reconstruction} 
\begin{figure}
\hspace{3.5cm}
\includegraphics[scale=0.4]{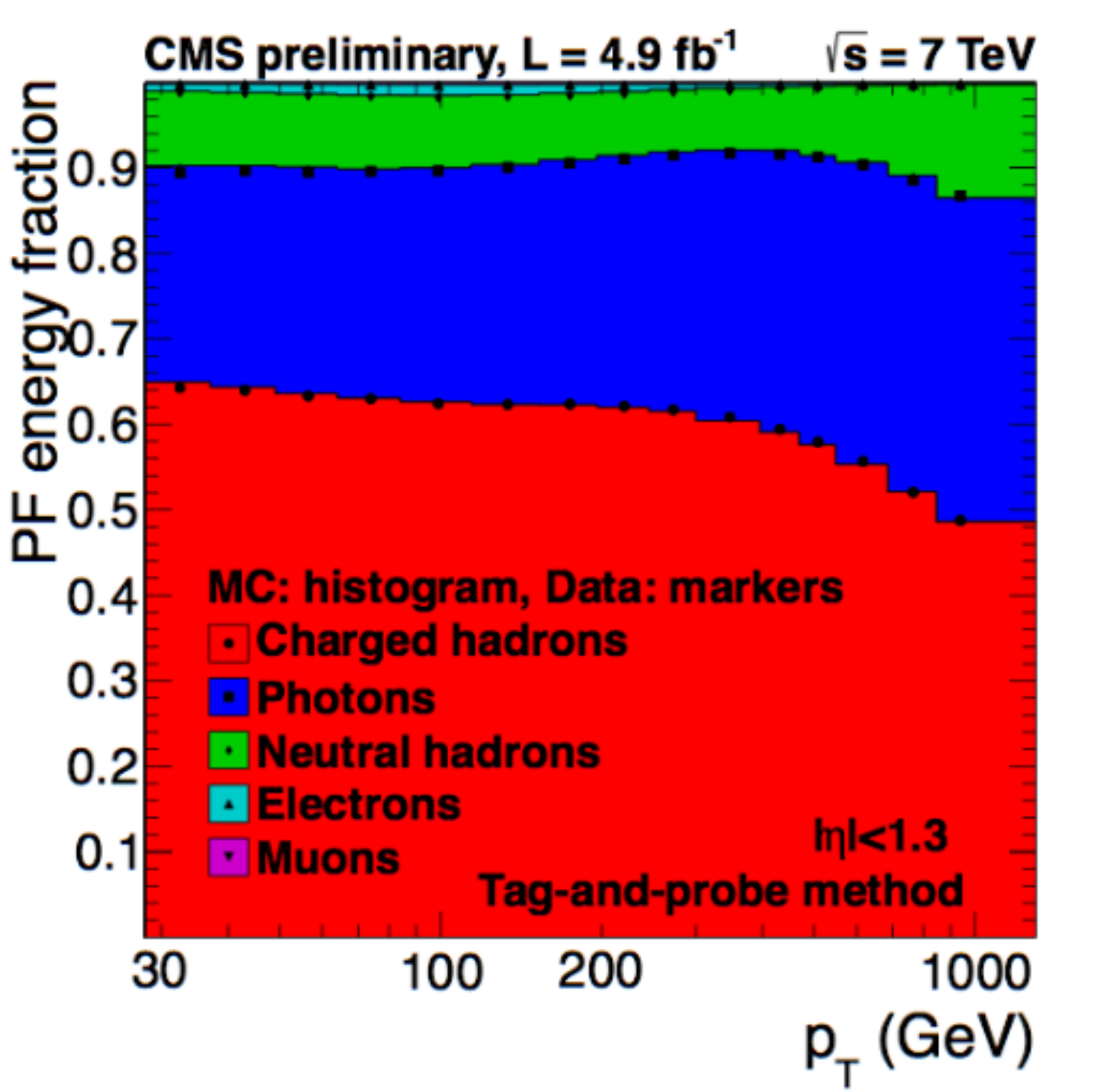}
\caption{Jet compositions}
\label{fig:compositions}
\end{figure}

 In this presentation we have discussed about jet, which are reconstructed by particle flow (PF) algorithm. PF algorithm is an event
 reconstruction technique which attempts to reconstruct
 and identify all stable particles in an event. It combines all information from all the sub detectors viz. tracker, electromagnetic
 calorimeter (ECAL), hadronic calorimeter (HCAL), muon chamber. The stable particles reconstructed from these subdetectors are
 $\mu^{+},~~ \mu^{-},~~ e^{+},~~ e^{-},~~ \gamma,~~ \pi^{+},~~ \pi^{-},~~ \pi^{0},~~ K^{+},$ \newline
 $ K^{-},~~ K^{0}$ etc.
 Jet reconstruction algorithm is applied on this final state stable particles to
 reconstruct jets. Figure~\ref{fig:compositions} shows the data montecarlo comparison of jet energy fraction of the constituents. 
 In our current presentation, we have shown analysis with jets reconstructed by anti-kT algorithm, using size parameter R = 0.7
 (ak7).
 The clustering is done using FastJet package. 

 \section{Jet Energy Correction} 
\begin{figure}
\includegraphics[scale=0.35]{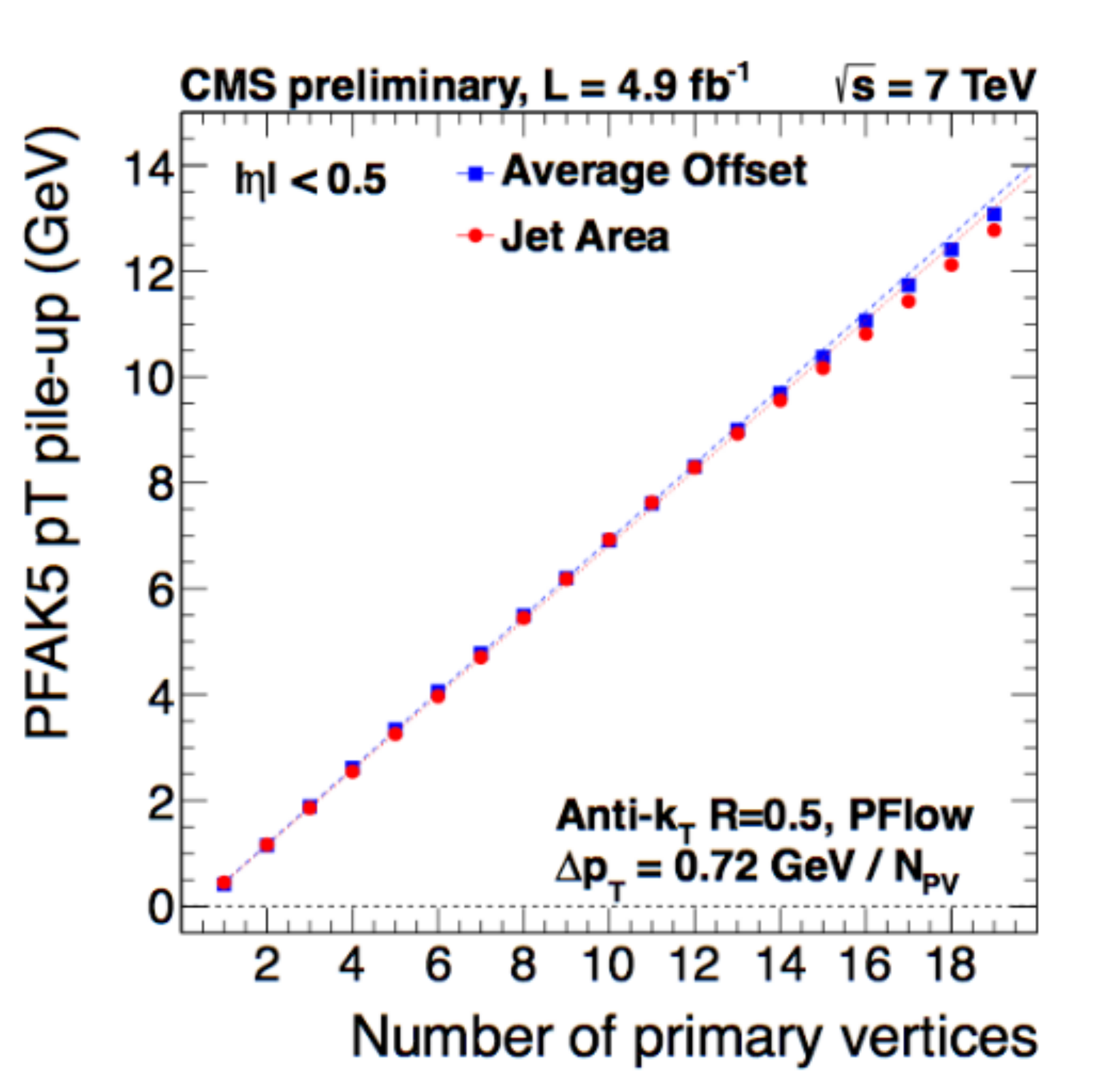}
\includegraphics[scale=0.35]{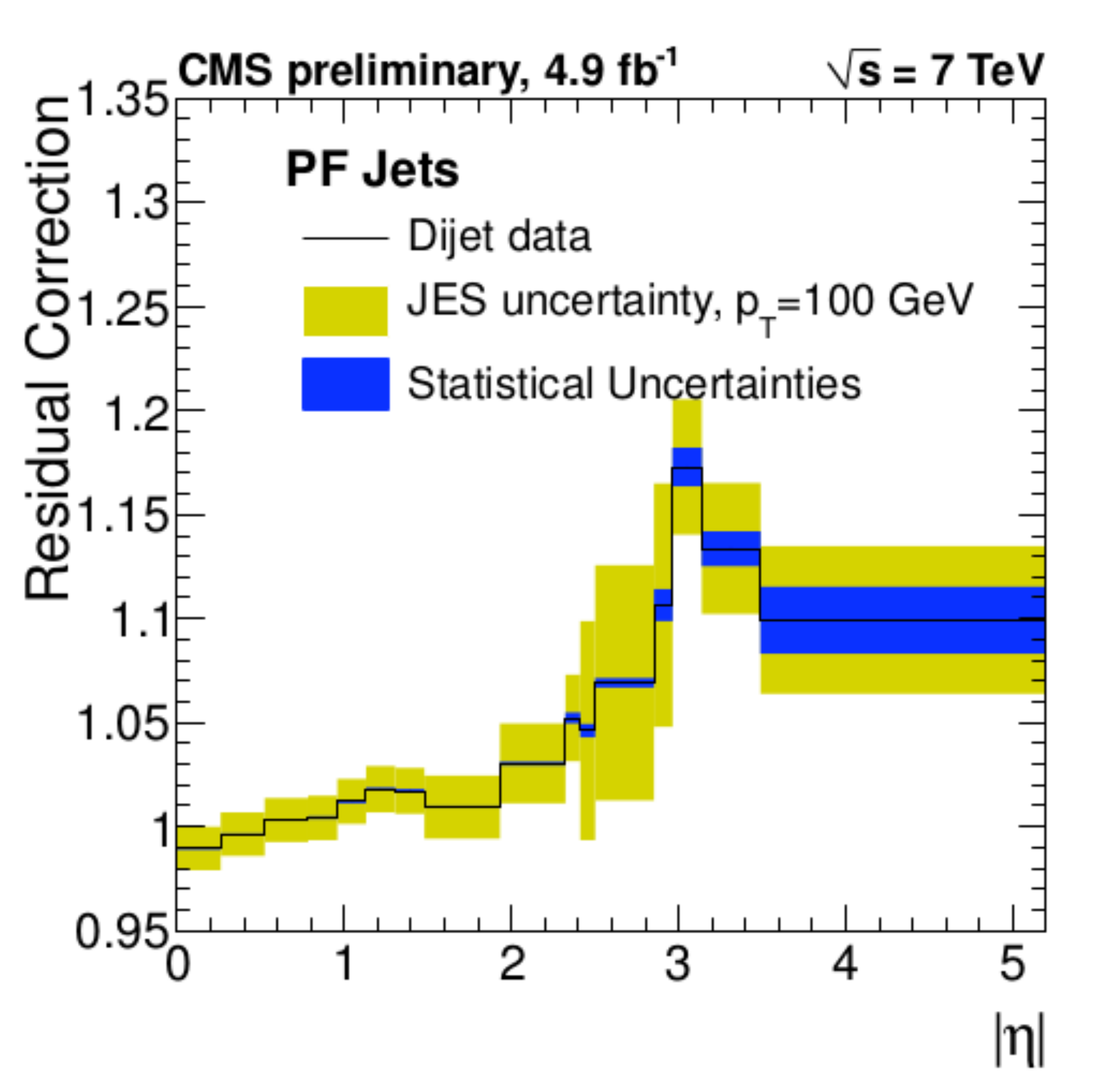}
\includegraphics[scale=0.45]{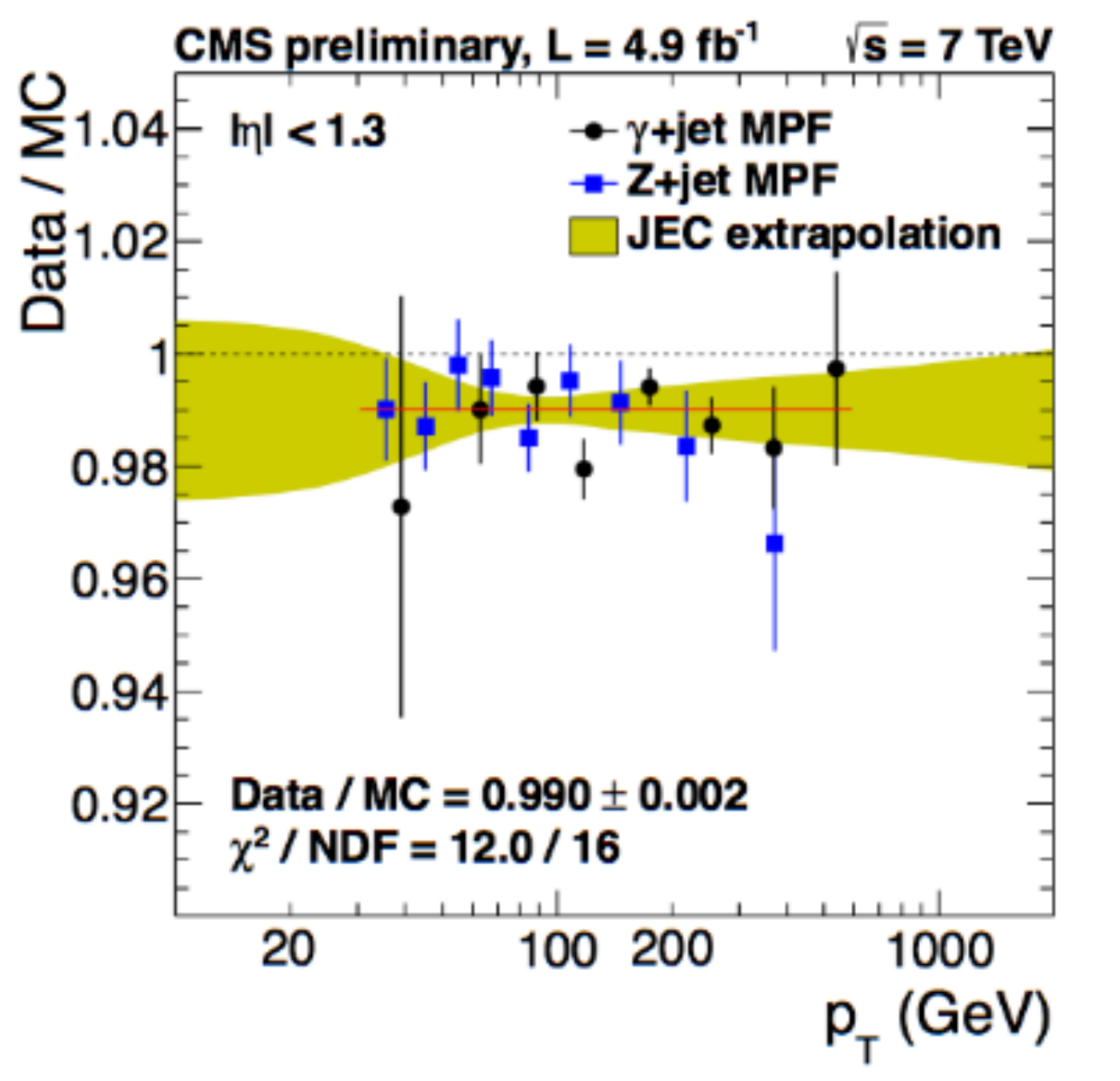} 
\hspace{2.1cm}
\includegraphics[scale=0.44]{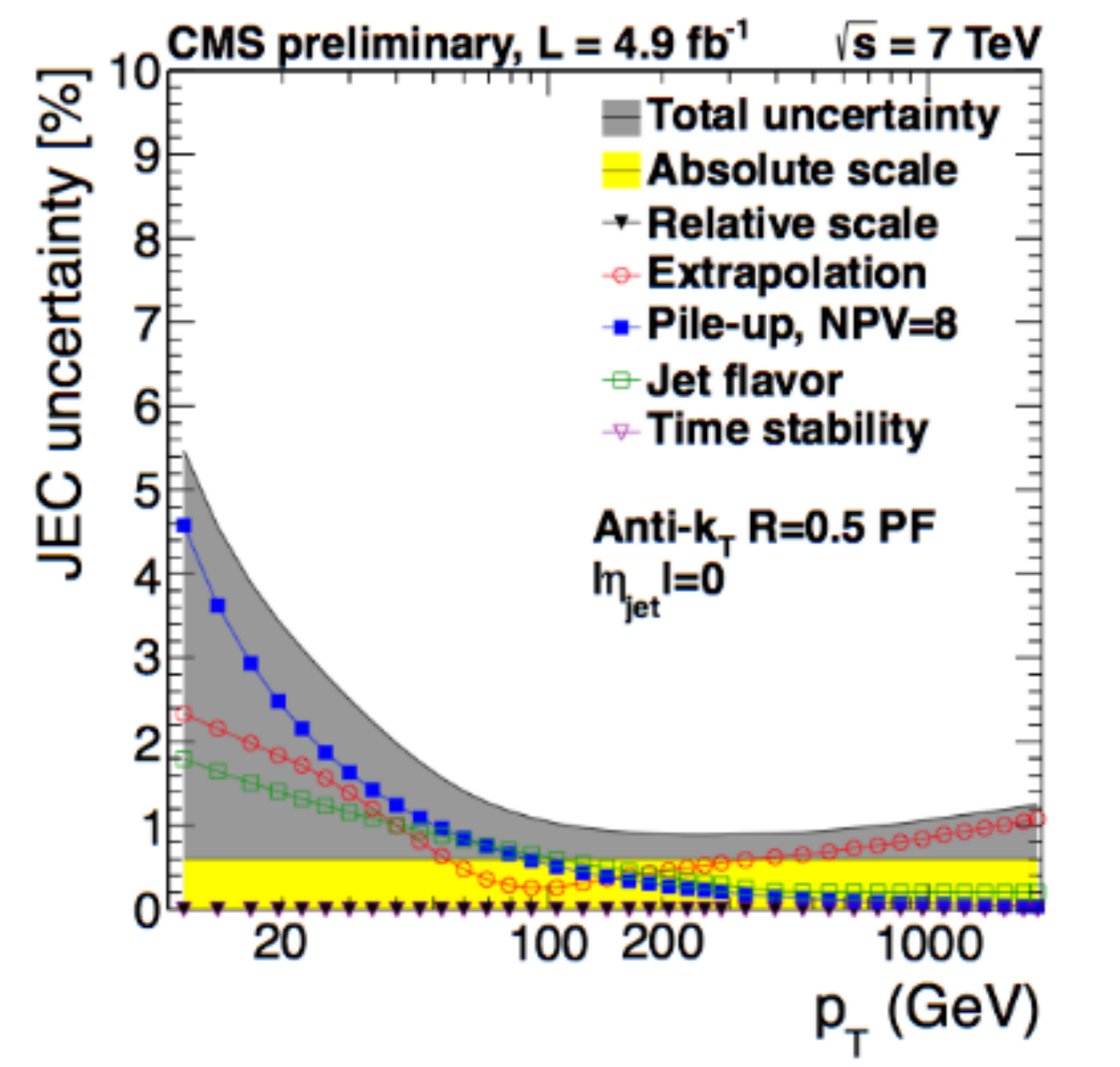}
\small
\caption{JEC factors and JES Uncertainties.}
\label{fig:jec}
\end{figure}

 Jets are reconstructed from stable particles in the detector. So jet energy and momenta are effected by detector level noise,
 high pileup, varying detector response etc. The purpose of jet energy correcion (JEC) is to relate, on average, the enrgy measured
 for the detector jet corresponding to the particle level jet.\cite{JES} A true particle level jet results from clustering
 of all stable
 particles originating from fragmentation of hard scattered partons, as well as particles orginating from underlying event
 activity. \newline In CMS, the JEC is applied on each momentum component of the reconstructed jet as a multiplicative factor as 
 $ p_{\mu}^{cor}~=~\mathcal{C}.p_{\mu}^{raw}$,
 where the correction factor $\mathcal{C}$ is product of several components as,  
 \begin{equation}
 \mathcal{C}~=~C_{1}(Offset)\times C_{2}(Relative) \times C_{3}(Absolute) \times C_{4}(Residual) 
 \end{equation} 
 Here $C_{1}(Offset)$ is the correction factor which removes the contribution coming from pile-up. $C_{2}(Relative)$ corrects for the
 non linear $\eta$(pseudo rapidity $\eta=-ln(tan \frac{\theta}{2})$) dependence of the detector response. $C_{3}(Absolute)$ for the
 non linear jet transverse momentum ($p_{T}$) dependence of the detector response and $C_{4}(Residual)$ is the residual correction
 factor which accounts for the mismatch between data and montecarlo prediction. \newline The $C_{1}(Offset)$ factor is measured by
 average offset method and jet area method. \newline $C_{2}(Relative)$ factor method is
 determined by dijet balance method and $C_{3}(Absolute)$ is factor
 is determined by $\gamma/Z+jet$ $p_{T}$ balance method. Figure~\ref{fig:jec} shows the individual JEC factors and uncertainty due
 to several factors which are uncorrelated among themselves. 
 
\section{Inclusive And Dijet Measurements}
 In this talk we have mainly discussed about the measurement of double differential cross section of inclusive jet
 ($p+p \rightarrow jet+X$) and dijet
 ($p+p \rightarrow jet+jet+X$) production as a function of inclusive jet $p_{T}$ and dijet invariant mass ($M_{jj}$) for different
 rapidity bins from $\sqrt{s}=$ 7 TeV CMS data\cite{QCD}. The measured cross sections are corrected for detector
 effects and compared
 to the QCD
 predictions. The parton transverse momentum fractions $x_T=\frac{2p_{T}}{\sqrt{s}}$ probed in this measurement cover the range
 $0.033<x_T<0.57$. \newline The jets are selected with tight jet identification criteria, to reject detector level noise,
 followed by JEC applied on selected jets. The double differential cross section are measured using the formula 
 \begin{equation} 
 \frac{d^2 \sigma }{dp_{T}d|y|} = \frac{1}{\epsilon L}\frac{N}{\Delta p_{T} \Delta |y|} \times C_{unsmearing} 
 \end{equation}
 \begin{equation}
 \frac{d^2 \sigma}{dM_{jj}d|y|} = \frac{1}{\epsilon L}\frac{N}{\Delta M_{jj} \Delta |y|} \times C_{unsmearing}
 \end{equation}
 where $L$ is the total integrated luminosity, $\epsilon$ is the trigger efficiency, $N$ is the number of jets in a bin,
 where $\Delta p_{T}$, $\Delta M_{jj}$ are the bin width and $C_{unsmearing}$ is the non perturbative correction factor. \newline
  For the inclusive
 jet measurement, events are required to contain at least one tight jet with $p_{T}>$114 GeV, 196 GeV, 300 GeV,
 362 GeV, and 507 GeV for the five single-jet HLT triggers used respectively. For the dijet measurement, at least two tight
 reconstructed jets with ${p_{T}}_1>$60 GeV and ${p_{T}}_2>$ 30 GeV are required.
 
\begin{figure}
\includegraphics[scale=0.33]{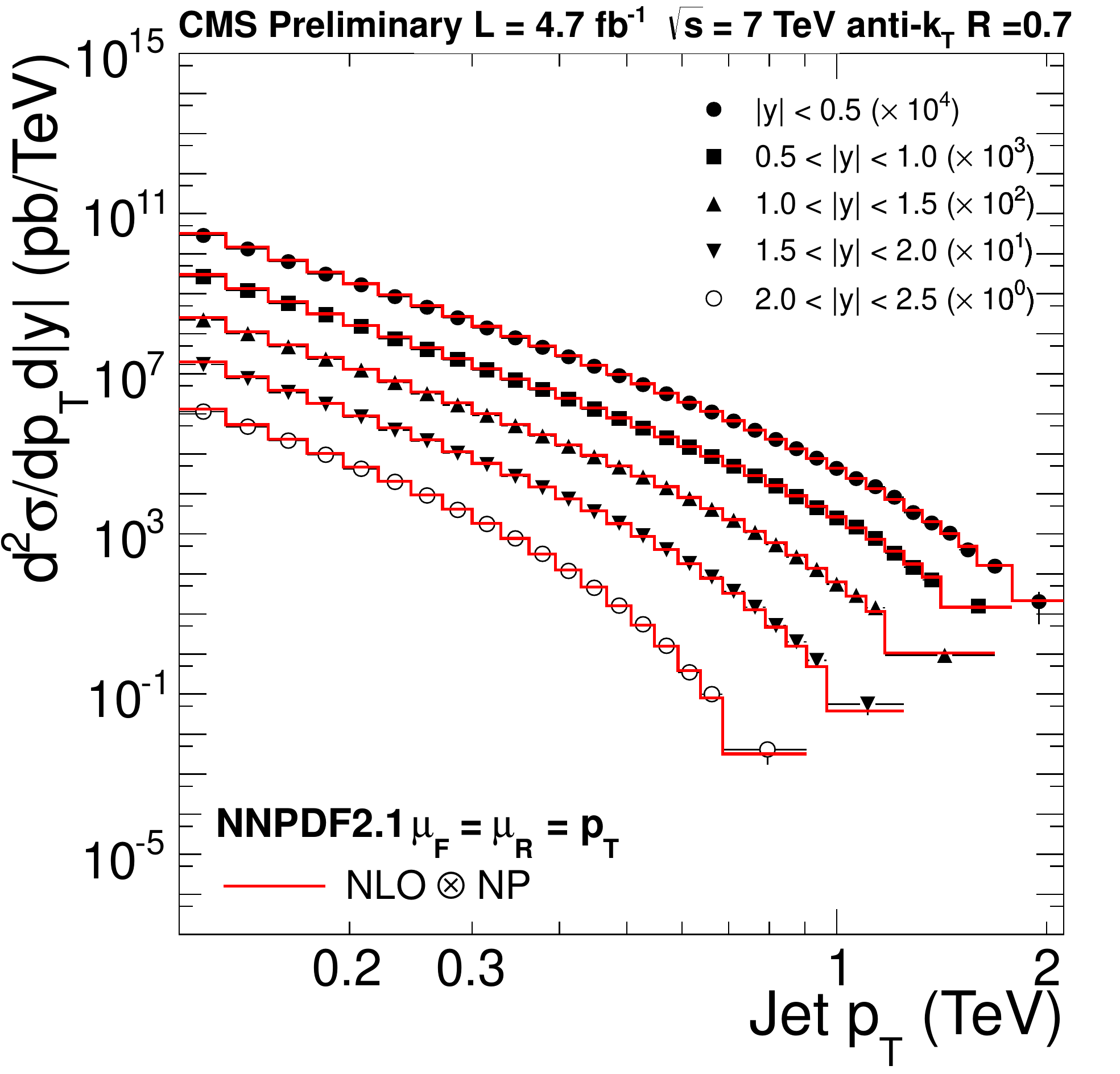}
\hspace{1cm}
\includegraphics[scale=0.33]{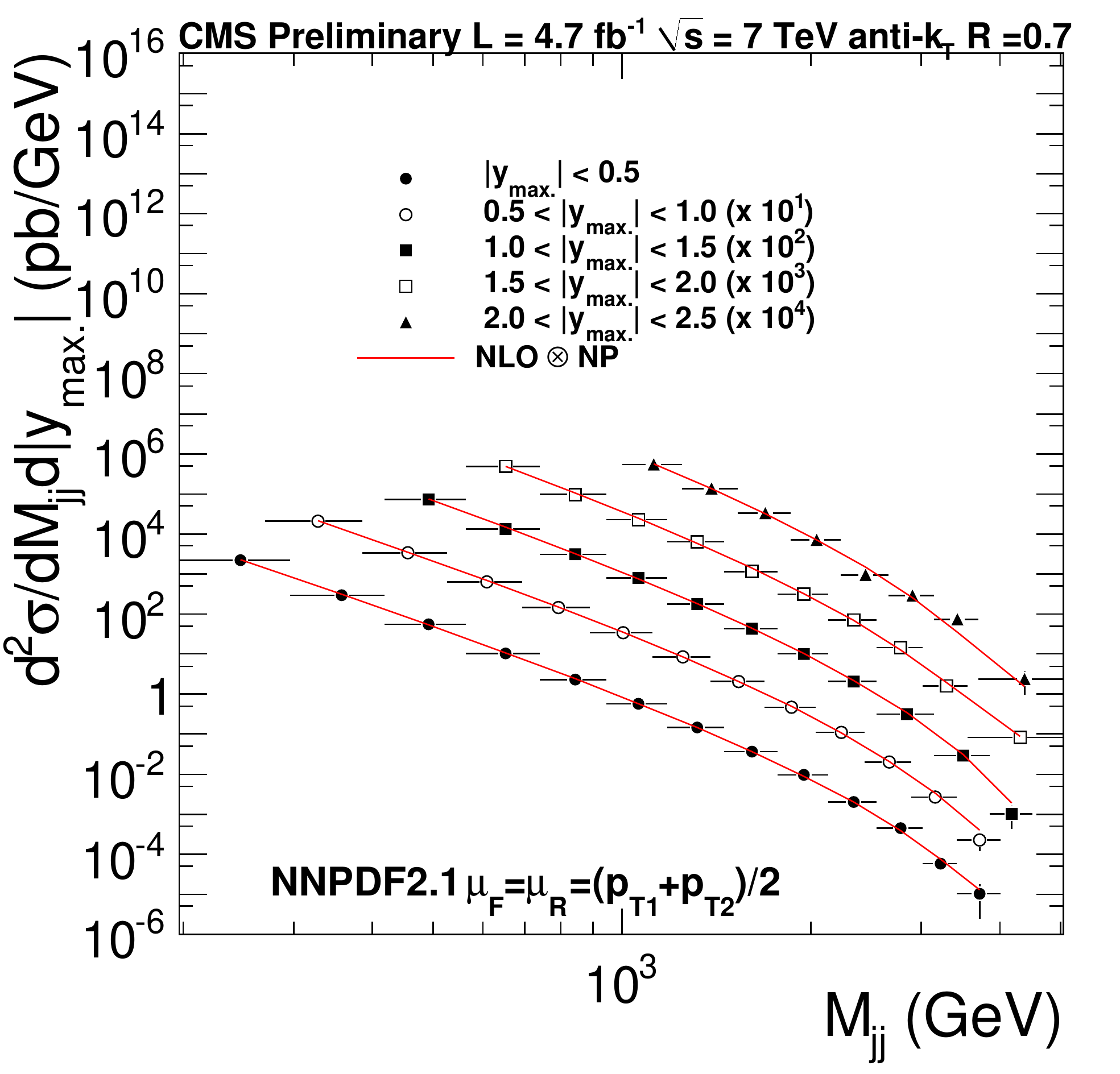} 
\caption{Inclusive and Dijet double differential cross section compared with $NLO\times NP$ theory prediction}
\label{fig:spectrum}
\end{figure}

\begin{figure}
\includegraphics[scale=0.35]{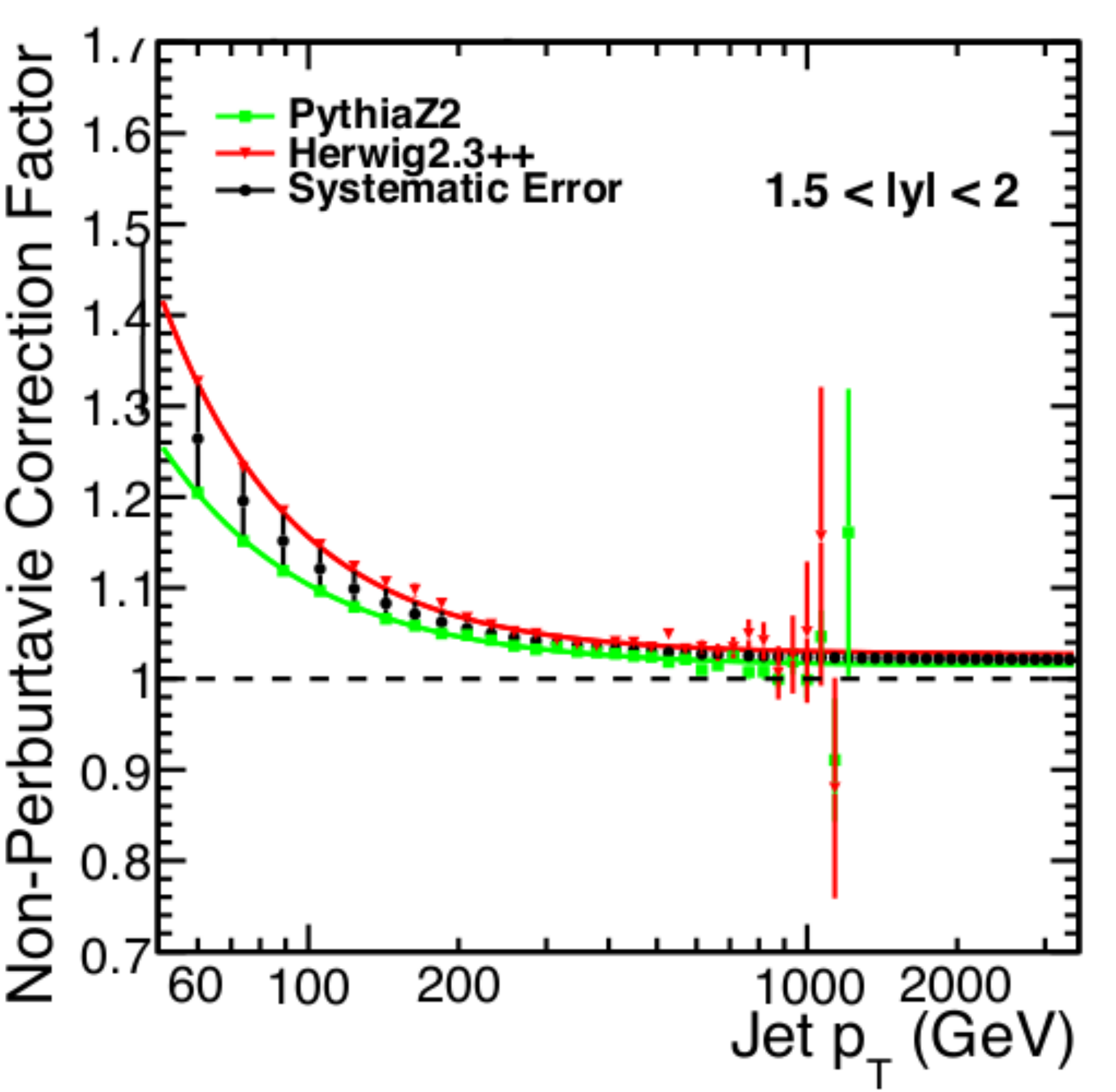}
\hspace{1.2cm}
\includegraphics[scale=0.34]{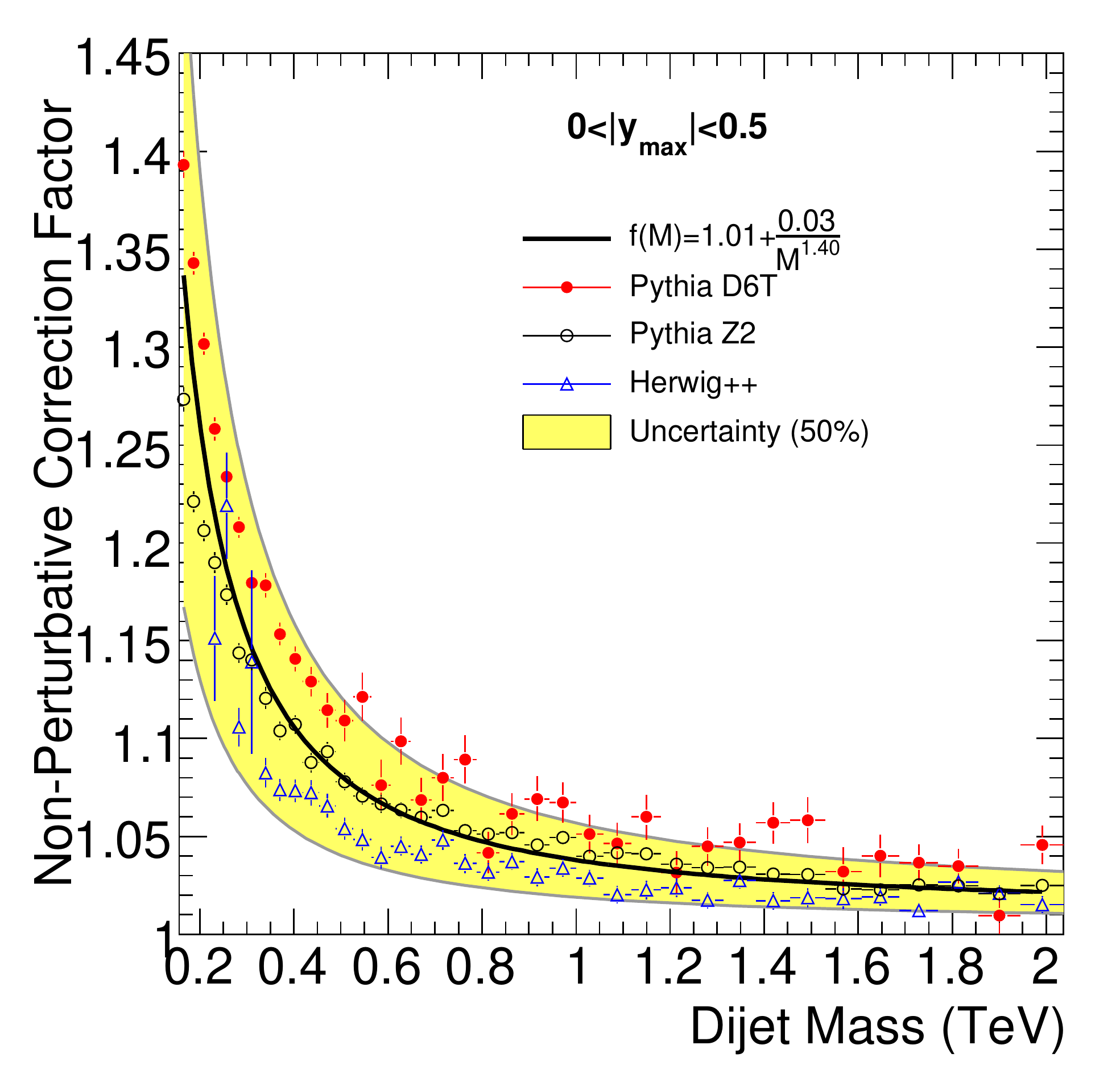}
\caption{NP correction factor for inclusive jet and dijet}
\label{fig:npcor}
\end{figure}

\begin{figure}
\includegraphics[scale=0.37]{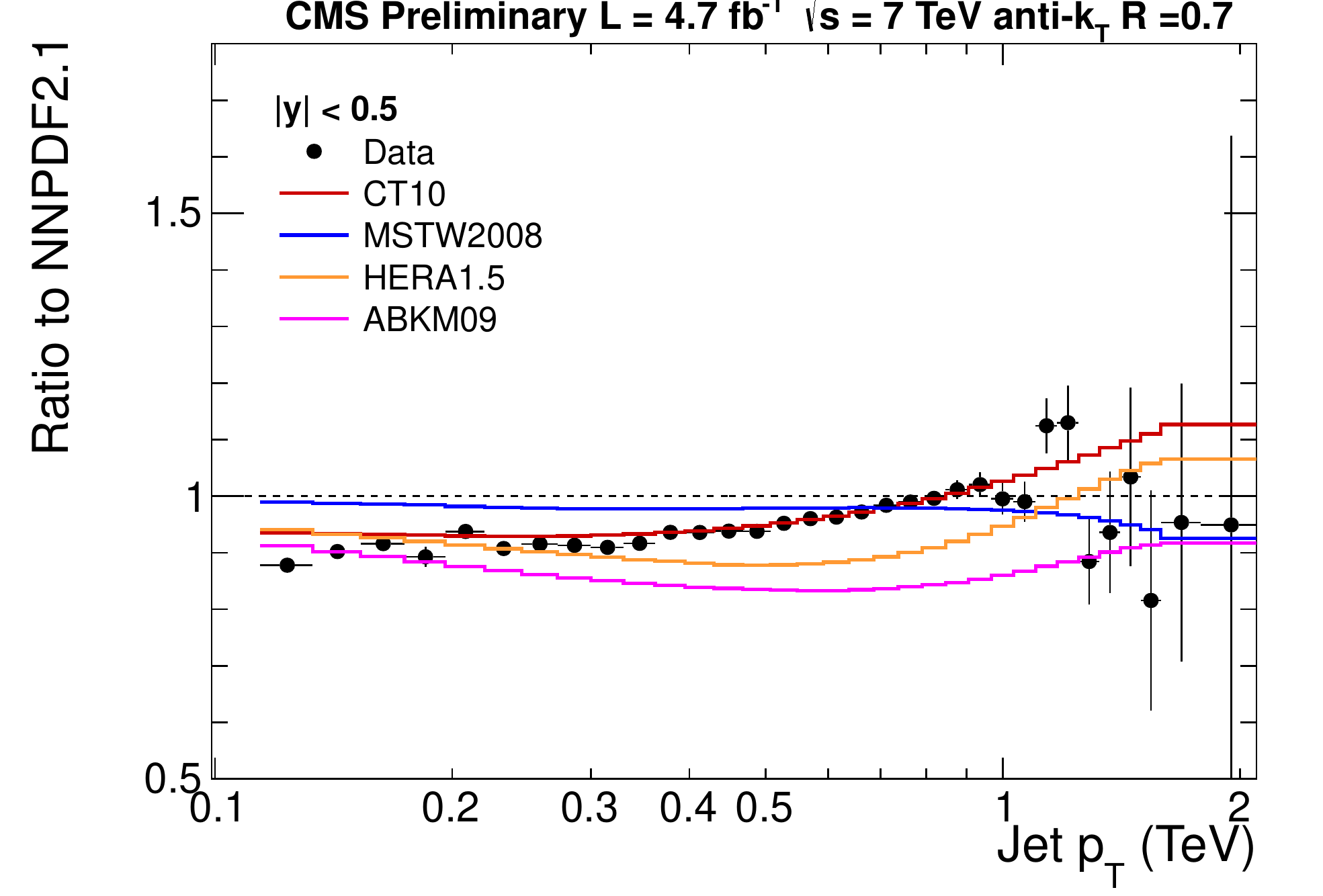}
\includegraphics[scale=0.37]{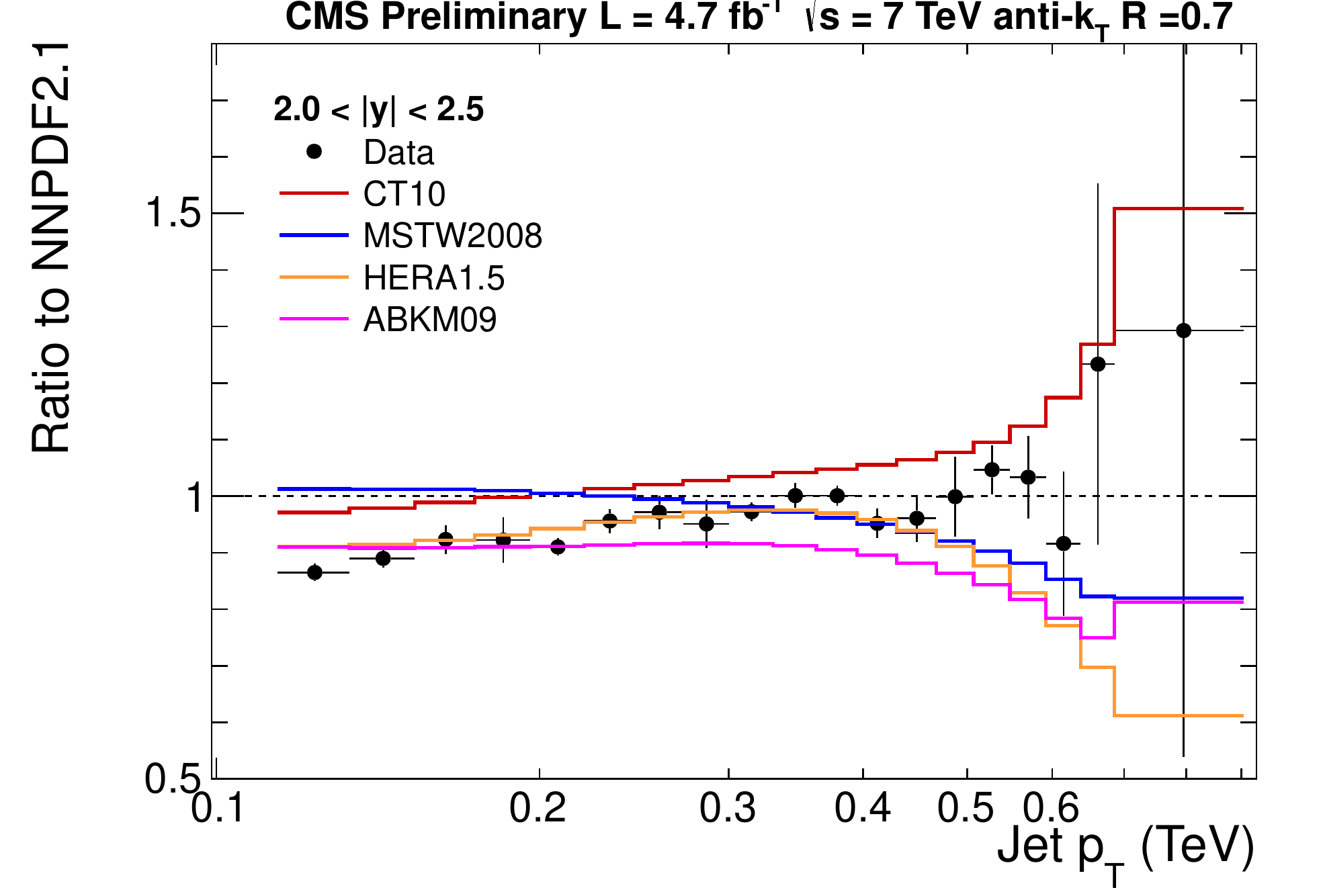}\\
\includegraphics[scale=0.37]{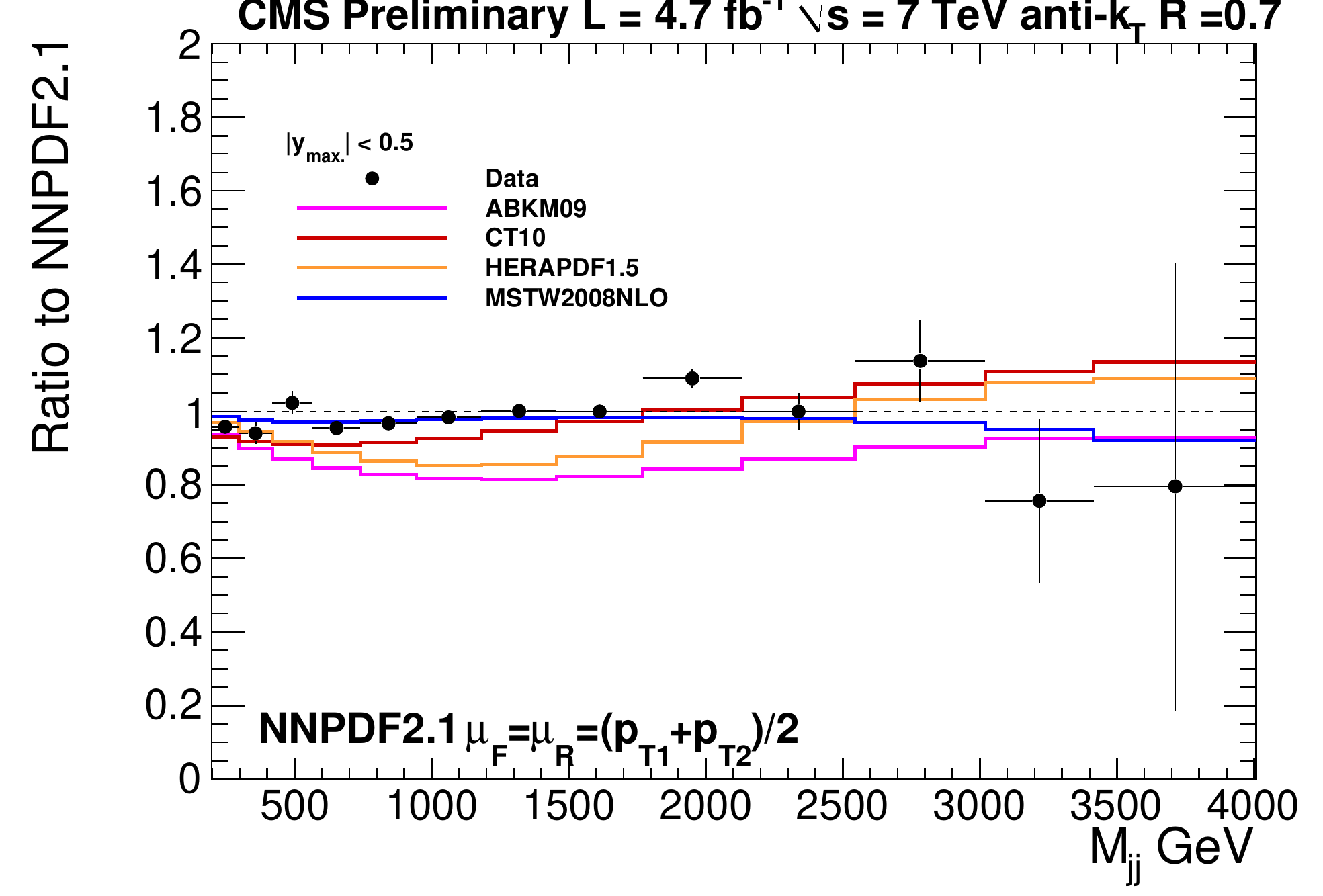}
\includegraphics[scale=0.37]{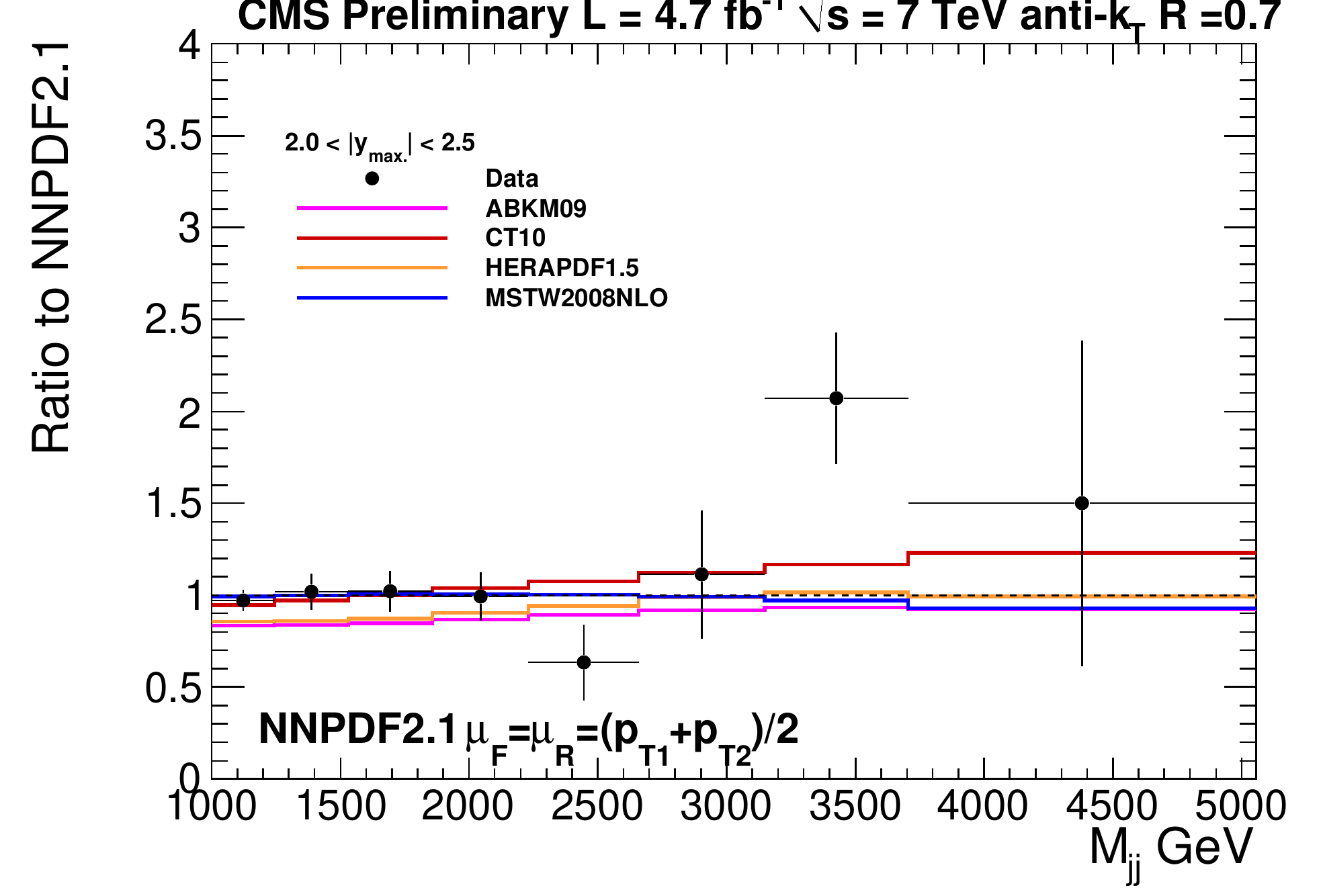}
\caption{Data over theory ratio with total uncertainty limits.
}
\label{fig:ratio}
\end{figure}

 Figure~\ref{fig:spectrum} shows the double differential spectrum for different rapidity bins extending upto $|y|=2.5$. The
 measured spectra is then unfolded using Bayesian unfolding (D'Agostini)\cite{ROOUNFOLD} technique to get rid of the detector resolution effects
 and then compare directly to the particle level spectrum. Comparison 
 with $NLO\times NP$ theory has been done for several parton distribution function (PDF) sets. \newline The non perturbative
 correction factor is applied on next to leading order (NLO) spectrum to account for hadronization effects and multi parton
 interactions (MPI). This factor is computed as the ratio of the observables, with once switching on and then switching off the
 hadronization and MPI effects. In the current analysis the NP currection factor is evaluated from two generators, Pythia6 and
 Herwig++. The difference in the factor derived from these two event generators is counted as the uncertainty introduced due to NP 
 correction on the measured cross section. Figure~\ref{fig:npcor} shows the variation of NP correction factor as a
 function of jet $p_{T}$ and
 $M_{jj}$. \newline The NLO theory calculation is done by NLOJet++ package using fastNLO program\cite{FNLO}, the renormalization
 and
 factorization scale has been set
 equal to inclusive jet $p_{T}$ and $M_{jj}$ in the individual cases. The scale uncertainty is evaluated by varying the scales for
 six points as $\frac{\mu_{F}}{\mu},\frac{\mu_{R}}{\mu}=(2,2),(0.5,0.5),(1,0.5),(0.5,1),(2,1),(1,2)$. The total theoretical
 uncertainty varies upto $30 \%$, where as the experimental uncertainty varies between $10 \%$ to $30 \%$. The NLO theoretical
 calculations are done for five different PDF sets viz. ABKM09, MSTW2008, HERA15, CT10, NNPDF2.1 . Figure~\ref{fig:ratio} shows
 data over theory plots for both the variables in central and outer most rapidity bin. The figure shows that the measured spectrum
 is in good
 agreement with the NLO theory spectrum within experiemntal and theoretical uncertainty limits. 

\section{Conclusion}
 In this talk first we have discussed about the jet reconstruction and jet energy corrections and finally inclusive and dijet
 measurements from 7 TeV CMS data. A discussion on jet reconstruction from particle flow algorithm is done. Determination of
 individual JEC factors from data are discussed. Then we proceed to discuss about the double differential cross section
 measurement for
 inclusive and dijet production cross section from p-p collision. A comparison of unfolded measured cross section with
 $NLO\times NP$ theory spectrum is presented
 for five different PDF state for each rapidity bins. \newline Estimation of theoretical and experimental uncertainties have been
 discussed and it is shown that the data over theory ratio is well within the uncertainty bands. \newline This measurement can be
 used for PDF fittings and strong coupling constant extraction. 


\end{document}